# Gigawatt-hour Scale Savings on a Budget of Zero: Deep Reinforcement Learning based Optimal Control of Hot Water Systems

*Hussain Kazmi[1,2], Fahad Mehmood[3], Stefan Lodeweyckx[2], Johan Driesen[1]*

[1] ELECTA, ESAT, KU Leuven, Belgium
[2] Data Science, Enervalis, Belgium
[3] Suleiman Dawood School of Business, LUMS, Pakistan

**Abstract**

Energy consumption for hot water production is a major draw in high efficiency buildings. Optimizing this has typically been approached from a thermodynamics perspective, decoupled from occupant influence. Furthermore, optimization usually presupposes existence of a detailed dynamics model for the hot water system. These assumptions lead to suboptimal energy efficiency in the real world. In this paper, we present a novel reinforcement learning based methodology which optimizes hot water production. The proposed methodology is completely generalizable, and does not require an offline step or human domain knowledge to build a model for the hot water vessel or the heating element. Occupant preferences too are learnt on the fly. The proposed system is applied to a set of 32 houses in the Netherlands where it reduces energy consumption for hot water production by roughly 20% with no loss of occupant comfort. Extrapolating, this translates to absolute savings of roughly 200 kWh for a single household on an annual basis. This performance can be replicated to any domestic hot water system and optimization objective, given that the fairly minimal requirements on sensor data are met. With millions of hot water systems operational worldwide, the proposed framework has the potential to reduce energy consumption in existing and new systems on a multi Gigawatt-hour scale in the years to come.

## 1. Introduction

Growing realization of the devastating impact of climate change (IPCC , 2014) has accelerated investments in energy efficiency (IEA, 2016) and renewable energy sources (IEA, 2017), (IRENA, 2017). As proliferation levels rise, mass produced solutions can yield only diminishing returns and human-centric considerations will become more important (Masoso, 2010). Energy efficient buildings serve as a useful case for this. While high thermal losses from energy inefficient buildings can be reduced with façade and insulation improvements, occupant behaviour inevitably becomes the final frontier.

Energy demand in high efficiency buildings has been demonstrated to be a function of occupant behaviour (Majcen, 2016). This is especially true for the energy consumed for spatial heating and domestic hot provision in modern, nearly zero energy buildings (Santin, 2009), (Gill, 2010). The thermal comfort preferences of building occupants play a major role in determining how much energy is consumed to heat up a building. Even more striking is the case of domestic hot water provision which already accounts for over 10% of residential energy demand in many countries (Pérez-Lombard, 2008) and will take on increasing importance as building technology progresses. Domestic hot water requirement in a household is strongly correlated with occupant demographics and behaviour (Dane George, 2015). The higher the hot water consumption, the higher the energy required to meet this demand (Kazmi, 2016). Other factors influencing the energy consumption include the ambient conditions and thermal equipment installed in the building such as the storage vessel and the heating element type.

Reducing the energy consumed for hot water vessels is possible by adopting more efficient thermal equipment. This can be in the form of a more efficient heating mechanism, for example replacing electrical resistance heating by a heat pump (Chua, 2010). The second route to higher efficiency is via an improved storage vessel which has lower ambient thermodynamic losses, better stratification etc. Work in this direction has continued apace and modern hot water systems boast energy efficiency much higher than their counterparts from older generations.

However, most of these optimizations are carried out in design phase and have been grounded in thermodynamics. Consequently, they ignore the two most important components of the system: the operational aspect of the hot water system and the human user. The allure of such optimization should be evident: there are over 200 million European households alone. Even marginal increases in efficiency in hot water production for these households with their existing systems can have substantial economic and environmental impact.

Prior work in this direction has mostly revolved around improving performance of rule based controllers by employing the Model Predictive Control (MPC) family of algorithms (Oldewurtel, 2012), (Afram, 2014). MPC offers the flexibility to optimize towards multiple objectives such as energy efficiency, maximizing self-consumption of local renewable generation etc. However, MPC suffers from a major drawback for residential buildings: control presupposes a reliable model of the devices under consideration. This model is necessary for optimization and while it may be economically viable to develop detailed building models for large commercial buildings, it is seldom cost-effective to undertake the same effort for residential buildings. When such a model is not available, MPC is unable to perform optimal control. Additionally, even when a model is available, problems can arise because of the non-adaptive nature of the model which means operational changes to hardware or an inaccurate initial model can lead to sub-optimal performance and loss of occupant comfort.

A framework to integrate a data-driven model of the storage vessel with active control based on occupant preferences has been presented in (Kazmi, 2016). However, despite being data driven, the work is limited in the sense that it includes an offline learning step to model the storage vessel. While an improvement over traditional MPC, the work still can't be generalized to the huge diversity of hot water systems being used by individual households. In a similar vein, optimization has also been considered from the perspective of maximizing occupant comfort while minimizing ambient losses (Zhang, 2007). In this case again, the heat loss model is not data driven and assumes knowledge of physical properties of the vessel which is usually an unrealistic assumption in most existing systems.

Work on model-free reinforcement learning has also been proposed as a way to sidestep complications arising out of MPC's inflexibility and the requirements for a prior model (Frederik Ruelens, 2014) (Jose Vazquez-Canteli, 2017). However, in the former work, the objective investigated is not energy efficiency and a prior model of the storage vessel is still assumed. In the latter work, the optimization focuses on using a hot water storage vessel to provide warmth to a building and is therefore unaffected by the specific temperature distribution inside the storage vessel. Dynamic programing requiring little computational power onsite as all the computations are performed in advance has been proposed in the past to tackle this problem as well (F. De Ridder, 2011), (James, 2003). However these works suffer from the same generalizability problems as mentioned earlier.

In a similar vein, completely black box models to learn building behaviour and then using these for active control have been proposed (Frederik Ruelens S. I., 2015), (Tianshu Wei, 2017). However, the building envelop displays vastly different characteristics than a hot water vessel because of nonlinear dynamics, stronger stratification effects and the nature of interaction with end users.

In this paper we present, to the best of our knowledge for the first time, the possibility of both learning a storage vessel model online and controlling it towards a predefined objective. Learning a model for the storage vessel directly from sensor data is challenging because sensing capabilities are limited in most real world systems: the most common configuration is a single mid-point temperature sensor. Since the temperature distribution in a storage vessel is usually non-homogeneous and non-linear, extrapolating from a single temperature reading to an entire distribution is difficult. In this paper, we demonstrate a framework to achieve this in practical settings with realistic sensor data, and apply the proposed algorithm to a real world case study of optimizing hot water production in 32 net-zero energy buildings in the Netherlands.

Historically, there has always been a trade-off associated with white, grey and black-box models to model hot water systems. White box models, as defined in the literature, model the hot water system according to thermodynamics. Black box models on the other hand make use of sensors to learn system dynamics and grey box modelling techniques combine elements from the two (Afram, Review of modeling methods for HVAC systems, 2014). Previously black box models learnt online have not been used to optimize hot water systems for energy efficiency because of issues with learning storage vessel state as explained above. The proposed framework changes this by making use of a novel model-based reinforcement learning algorithm.

It is important to note that the developed framework can be complementary to previously developed optimal flow rate formulations (Badescu, Optimal control of flow in solar collectors for maximum exergy extraction, 2007), (Badescu, 2008). More specifically, the current research intends to answer the question of when to reheat the storage vessel based on dynamically learnt occupant consumption patterns and their impact on energy content in the storage vessel. This is not constrained by the internal operation of the heating element as long as it can provide the required power. In doing so, it assumes that the heating element operates in a binary (on-off) mode but continually makes internal decisions on optimal flow rates once switched on. Deeper integration of the two optimization problems has the potential to further improve energy efficiency and is one promising avenue for future research.

Improving energy efficiency of storage vessels in this way is extremely useful because it is applicable not only to new devices but also to the millions of devices already installed in households. Considering the fact that many households consume more than a megawatt-hour for hot water production annually, widespread adoption of the proposed schema has the potential to reduce annual global energy consumption by hundreds of gigawatt-hours, if not more.

Optimizing energy efficiency is not the only use case envisioned for the developed framework. Domestic hot water vessels and the accompanying heating elements have, over the years, acquired an interesting second life as a source of flexibility towards the energy grid (Koen Vanthournout, 2012). This will become progressively more important as distributed generation and smart grids concepts mature further. Using the hot water vessel as storage to provide flexibility to the grid can result in substantial cost savings from a grid reinforcement investment perspective. This is made more attractive still by the fact that hot water vessels can be used to provide flexibility throughout the year as opposed to the thermal mass of a building which is usually of a more seasonal nature. The proposed framework has the capability to perform this task seamlessly as well.

In the following, we describe the methodology followed to develop a reinforcement learning framework to achieve this. We present results for both the learning (of the models) and the control aspect. To validate the online learning component, we present results from both offline and online tests. Likewise, we demonstrate that performing optimal control using this learnt model results in energy savings approaching 20% in real houses without compromising end user comfort. Finally, we conclude with some reflections on extending the proposed framework to integrate distributed generation and provide ancillary services to the electric grid.

## 2. Methodology

The research question at the heart of this paper is whether it is possible to further improve performance of thermodynamically optimized hot water systems by taking into account occupant interaction and hot water system dynamics. This optimization is made possible according to the observations in (Kazmi, 2016) (Zhang, 2007): (1) postponing demand in time to when it is actually required by the end user can reduce thermodynamic losses, and (2) reheating when the ambient temperature is more amenable to the heating process can improve efficiency. This latter concern is only true for technologies such as heat pumps while the former holds true for most heating elements.

The obvious additional constraint when compared with state of the art is that no prior information about the hot water system is available and therefore needs to be learnt on the fly as well.

In this paper, we propose a data driven approach to optimizing any residential hot water system limiting ourselves to only the sensors available by default. These include an energy meter for the heating element, a flow meter for hot water consumption and a single temperature sensor mounted at the mid-point in the vessel. We do this to focus on the generalization potential of the proposed methodology rather than obtaining the theoretically best possible results. While increasing the amount of sensors would facilitate learning of a more accurate model, it also increases costs. Starting from no prior information about the hot water system or the occupant, the reinforcement agent learns to optimize the energy efficiency of the vessel. With time and more data, the performance improves.

More specifically, without loss of generalization, we consider the case of a 200 litre domestic hot water vessel coupled with an air source heat pump in 32 net zero energy buildings. Prior to active control, everything about the thermal system and its environment is assumed to be unknown.

There are multiple dimensions to this problem then. The first one being to learn an accurate representation of the different elements involved: the storage vessel, the heating element and the human users of hot water. The second one is to use these representations to help improve the energy efficiency of the system. This two-step approach is in contrast to the single step model-free techniques highlighted earlier (e.g. Q-learning, SARSA FQI etc.) in which optimal control actions are identified directly from system states and an explicit dynamics model isn't learnt.

### 2.1. Problem formulation

We begin by formulating the problem as a Partially Observable Markov Decision Process (POMDP): $\{S, A, T, R, O, \Omega\}$. The controller is a reinforcement learning agent since it has the joint task of learning a model of the thermal system while simultaneously optimizing towards some specified objectives. The thermal system is characterized by a certain state, $s \in S$, which can be affected by the reinforcement learning agent through a control action, $a \in A$ resulting in next state of the system $s_{t+1} \in S$ given by the function $T(s, a, \varepsilon)$:

$$s_{t+1} = f(s_t, a_t) + \varepsilon \qquad (1)$$

The transition function is assumed to be stochastic (indicated by the presence of $\varepsilon$), implying that given the same starting state and control action, the final state can vary according to some probability distribution, $\varepsilon \sim N(\mu, \sigma^2)$. The state transition results in the agent receiving a reward based on the initial and terminal states as well as the action taken, given by $R(s, a, s')$. Since we work with finite time horizon control, no temporal discounting is assumed resulting in nonstationary policies. The Markovian assumption here makes explicit the fact that the state of the thermal system depends entirely on the current system state and not previous ones. We further develop these components for the specific case of hot water systems:

**State and observations of state, s**: the state vector comprises of three individual components: the state of the storage vessel, the state of the heating mechanism and the state of the environment:

1. We define the **state of the vessel** as the embodied energy content. Since this energy content is not directly observable, observed temperature is used as a proxy. However, due to inadequate sensing (in this case, we assume a mid-point sensor but the framework is applicable regardless of the location of a sensor as long as it is exists), the problem becomes one of partial observability. In particular, a mid-point temperature observation fails to provide information about stratification and non-linear dynamics in the vessel, both key aspects in optimization.
2. The **state of the heating mechanism** is usually observable, although discrete sampling means this information is only available periodically and with some delay, usually on the order of minutes (5 to 15). For the purpose of DHW production, the heat pump under consideration can be in one of two states; an idle mode and an operational mode in which the heat pump reheats the storage vessel.
3. The **state of the environment** includes an indication about current and future occupant behaviour as well as other uncontrollable environmental factors such as the ambient temperature. There is an inherent uncertainty associated with these predictions because of stochastic human behaviour given by $o \in O$. The ambient temperature measurements and predictions are gathered through freely available resources on the internet (Weather Underground, 2017).

**Action, a**: the action vector is a periodically updated sequence of control actions $a_{0:T}$ (from the present moment until the planning time horizon, T). The action to be taken now by the heating mechanism is, in its simplest form, a binary decision, i.e. whether to reheat the storage vessel or not. The target temperature of the heat pump can also be controlled meaning the policy containing all control actions is a $[T \times 2]$ matrix.

**Transition function**: the transition function maps a given input state and control action to a future state. By chaining previous predictions of the transition function as input states, arbitrarily long predictions can be made for the future evolution of the hot water system. It is desirable for a transition function to predict future states accurately. Where this is not possible, it is desirable for the reinforcement agent to be cognizant of its limitations. This is made possible by learning a stochastic transition function along the lines of the popular PILCO (Deisenroth, 2011), TEXPLORE (Hester, 2013) and Deep PILCO (Gal, 2016) algorithms. TEXPLORE uses random forests to derive an estimate of the state uncertainty while Deep PILCO uses deep Bayesian neural networks for the same purpose. Following from this, we employ an ensemble of deep neural networks as the transition function approximation technique. The mean prediction of the stochastic transition function gives an indication of the likeliest next state, but the variance over the prediction explains how certain the agent is that this transition will materialize. In this case, the uncertainty in the vessel, heating element and environmental models stems from the following:

1. **Systemic uncertainty**: the reinforcement agent is less certain when making a prediction for regions in the state-space that are intrinsically noisy
2. **Sample uncertainty**: the reinforcement agent is less certain when generalizing to regions of the state-space it has not experienced before

Incorporating uncertainty in the decision making process allows for risk-averse decision making, e.g. the agent should decide to reheat when there is a small but non-negligible probability of there being insufficient hot water.

**Reward function**: the immediate (one-step) reward the agent receives at any time step is a function of the initial and final system states and the control action taken by the agent. These different components are necessary to distinguish between 'good' and 'bad' agent control actions. For our purposes, this reward stream is composed of multiple components:

1. **Lost occupant comfort, $c_t$**: in any human-centric optimization scheme, occupant comfort is of paramount importance. If user demand is met by water of at least 45°C then there is assumed to be no lost comfort. The amount of hot water consumed below this threshold defines $c_t$. The agent places a value of ∞ on occupant comfort which means all policies which lead to loss of comfort are infeasible. However, due to inaccuracies in occupant behaviour prediction means that there can still be some loss in practice.
2. **Optimization objective, $o_t$**: Since the goal of the present study is energy efficiency, the agent receives reward inversely proportional to its energy consumption. The more frugal an agent is with its energy consumption, the higher its reward.
3. **Exploration bonus, $e_t$**: this refers to the reinforcement agent taking exploratory steps in the state-action space for the sole purpose of improving its model of the system dynamics. Thus, every time the agent improves its model, it gets a reward proportional to the amount of information gain.

The overall reward, $r_t$, can be defined as:

$$r_t = A.c_t + B.o_t + C.e_t \qquad (2)$$

Where $A, B$ and $C$ are the individual weights given to occupant comfort, energy efficiency and the exploration bonus respectively. By valuing occupant comfort as infinite, the agent only considers policies which are as energy efficient as possible while still occasionally taking exploratory steps. Practically, this means $A$ has a very large value leading to the agent receiving a very high penalty (negative reward) every time user comfort is violated. Both $B$ and $C$ are valued similarly. In fact, the controller is quite insensitive to the exact value because, over time, the exploration bonus term decreases in relative importance as explained next.

**The exploration / exploitation dilemma**: One of the central challenges in reinforcement learning is the exploration / exploitation dilemma. This refers to balancing the exploration bonus highlighted earlier against other objectives (occupant comfort and operational efficiency). Since we assume absolutely no prior information about the thermal system, every new observation helps the reinforcement learner gain valuable insights into system operation. This is true especially at the start of the learning process and is formalized by the notion of information gain, $H$ (Balian, 2004):

$$H = -\sum_{i=1}^{n} p(x_i) \log_b p(x_i) \qquad (3)$$

Information gain is considered high when an unlikely event happens or when the prior distribution of the agent is very uninformative. This effectively captures the sources of noise pointed out earlier. As the agent uses an ensemble of deep neural networks, there are $n$ neural networks learning the state transition function $y(x; w)$ such that the output $y$ is a nonlinear function of the input $x$ parametrized by weight, $w$. When a majority of the neural networks 'disagree' on the output $y$, the uncertainty is said to be high. Where this is caused by sample uncertainty, the agent has the opportunity to improve its representation by using the current sample in its learning process. Likewise, if the agent's uncertainty was low but the actual observation falls outside the prediction intervals, the agent should update its representation. The variance of this ensemble output is then used to drive exploration.

As the function approximation step gathers more samples in the state-action space, it improves its predictions and the variance collapses to the stochasticity inherent in the system (systemic uncertainty). The exploration bonus that the agent can derive from this variance therefore naturally decays over time as the agent gains more experience and grows increasingly confident in its predictions. Seeing multiple instances of the same experience conveys little additional information and the effect on the overall learnt model asymptotes eventually. This means the $C.e_t$ term diminishes in importance in the reward stream in the long term.

**The learning / planning dilemma**: once it is possible to map out future trajectories and their associated rewards using the learnt models, the next step is to maximize the reward stream. Completely enumerating all possible trajectories given even a deterministic scenario requires an exponential number of evaluations. More specifically, $2^T$ policies need to be computed given a time horizon $T$ and a binary decision variable at each time step. For real time control and a decision variable that is not strictly binary but also includes the target reheat temperature, it is infeasible to solve this problem exactly.

At the opposite end of the spectrum, lie the default strategies employed in most heating systems. These default strategies, in the face of such complexity, make use of simple rule based controllers that rely only on the most rudimentary state measurements e.g. reheating when the temperature falls below a threshold or at a fixed hour of the day. In the considered system, this policy takes the following form:

$$a_t = \begin{cases} 1, & if\ T_s < T_{tg} - \Delta T \\ 0, & if\ T_s \geq T_{tg} \end{cases} \tag{4}$$

Where $T_s$ is the temperature measured by the sensor and $T_{tg} - \Delta T$ is the temperature threshold which forces a reheat cycle, usually set between 45 and 50°C for residential users.

Between these two extremes of complete enumeration and constructing a single greedy solution lies the trade-off between computational efficiency and accuracy. We pose this problem as:

$$\max(R)$$

$$s.t.$$

$$HW_t^{\pi} \geq HW_t^{\pi_d},\ \forall\ k = \{0, 1, \ldots T\},\ HW_t^{\pi_d} > 0 \tag{5}$$

Where $HW_t^{\pi}$ and $HW_t^{\pi_d}$ are the amount of hot water in the vessel at time t following policy $\pi$ and the default policy $\pi_d$. The reward stream to be maximized is given as:

$$R(\pi\ |x_0) = \frac{1}{T}\sum_{t=t_0}^{t_0+T} r_t \tag{6}$$

Where $r_t$ is as defined in eq.2. This reward stream balances occupant comfort, energy efficiency and improvement of learnt models for the hot water system. The optimization problem itself can be solved by an exact method such as branch and bound (Lawler, 1966), a metaheuristic such as genetic algorithms (Srinivas) or ant colony optimization (Dorigo, 2006), a simpler heuristic or a combination thereof (Kazmi H. e., 2016).

In real time systems, the agent has to continuously update its models with recent observations besides planning its future actions using these models. Since training an ensemble of deep neural networks requires substantial computational power, this can only happen parsimoniously. We make the model calibration a function of the exploration bonus the agent has received. This leads to more frequent model updates when the agent gathers new experiences and vice versa. These concepts are summarized in Fig.1.

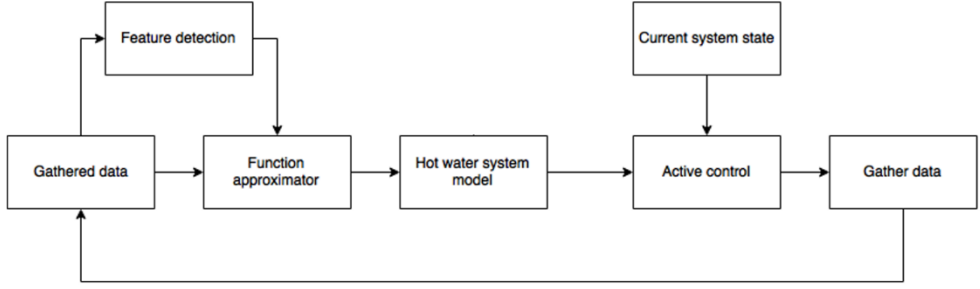

Fig. 1. Overview of proposed optimization system; the function approximation step refers to the discussed ensemble of neural networks while the active control component executes the planning strategy of the reinforcement agent

For completeness, we also present the proposed methodology in pseudocode form in Algorithm 1.

1. Initialize POMDP: $M \leftarrow (S, A, T, R, O, \Omega)$, $\pi \leftarrow \pi_d$, $\Delta H_T \leftarrow \infty$, $\Delta H_R \leftarrow \infty$
2. Initialize $H_{T,th}$, $H_{R,th}$
3. Repeat forever:
   (a) Update experiences: $s_t, a_t, r_{t+1}, s_{t+1}, \ldots$
   (b) $\Delta H_T \leftarrow var(T_{ensemble,i})$; $\Delta H_R \leftarrow var(R_{ensemble,i})$
   (c) If $\Delta H_T > H_{T,th}$
      i. Update the transition function, $T : S \times A \times S \rightarrow \mathbb{R}^2$
   (d) If $\Delta H_R > H_{R,th}$
      i. Update the reward function, $R : S \times A \times S \rightarrow \mathbb{R}^2$
   (e) Predict $O$, the occupant behaviour
   (f) Predict $T_a$, the ambient temperature
   (g) Rollout future trajectories, given $s_t$
   (h) Execute $\pi \leftarrow argmax_a(\mathbb{E}(r_t))$

Algorithm 1: Pseudocode describing the proposed methodology

**Results**

In the following, we describe the results obtained as a result of applying the proposed algorithm to a set of real houses in the Netherlands.

**Design of experiment**

For active control, we have split the set of 32 available houses into a default group of 13 houses (where a rule-based controller as described in eq. 4 is deployed) and an efficiency group of 19 houses (where the proposed framework is implemented). The 19 efficiency houses include a special set of 5 houses which have an extra temperature sensor installed at the outflow of the vessel to monitor consumed water temperature and evaluate loss of occupant comfort. This additional sensing was not used to facilitate active control in any way and so imparts no advantage to the proposed framework.

All of these houses have exactly the same hardware specifications for the heat pump and storage vessel. Moreover they are located in the same geographical area which means that weather does not affect efficiency of the different control groups. Since data about the demographics of households in the different groups was not available, the houses were split randomly into the two sets. Moreover, since the novelty of the proposed work is in determining an accurate storage vessel model on the fly, we validated the hot water model learnt by the reinforcement learner in both offline and online settings:

1. In the **offline validation test**, model predictions were compared against known outputs in an in-house lab setup. This allowed for detailed comparison of temperature distribution as predicted by the model vs. actual sensor measurements.
2. In the **online validation test**, model predictions were compared against temperature data readings coming from the mid-point sensor in real houses. This allowed for operational validation of predictive accuracy at the mid-way point, but did not allow for comparison between the entire temperature distributions.

In the following, we present detailed results for this validation as well as the efficacy of using this vessel model to optimize energy efficiency.

**Learning the thermal system characteristics**

1. Storage vessel model

The transition model for the storage vessel predicts the response of the storage vessel given a certain disturbance (such as passage of time or hot water draw). These ambient and mixing losses as a function of time and user consumption have to be learnt using the single mid-point sensor. Fig. 2 illustrates the model's representation for a vessel state given water flow of 100 litres after it has been reheated to 50°C. The three different curves highlight different ambient loss regimes (with the vessel left idle for 0, 12 and 24 hours). This representation has been learnt after a training period of four weeks with data from the actively controlled houses.

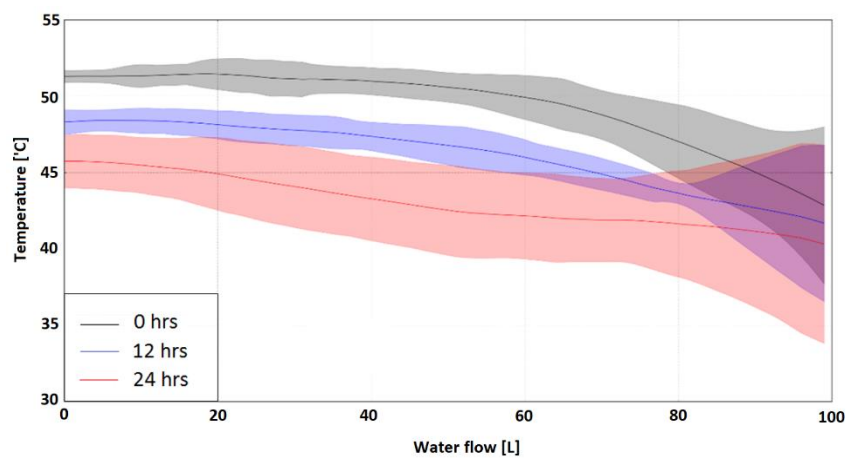

Fig. 2. Stratification and thermodynamic losses, as learnt by the vessel model

Variation in temperature along the x-axis shows the temperature drop with increasing water consumption from the storage vessel at the mid-point sensor. These are predictions from the learnt model and not sensor observations. The reduction in the starting temperature for the curves along the y-axis (red and blue compared with the black curve) identify the thermodynamic losses corresponding to an idle period of 12 and 24 hours. The shaded regions correspond to the uncertainty in the model's prediction. Uncertainty is lowest for cases of low flow (i.e. uncertainty increases with consumption) and for less delay since the last reheat cycle, i.e. uncertainty increases as the time between consumption and reheat cycle increases.

The goal for a reinforcement learner is to reduce this uncertainty over time by making more accurate predictions. This reduction in uncertainty over time is explored further in Fig. 3. As explained earlier, it is calculated as the variance over the ensemble input given the same starting conditions.

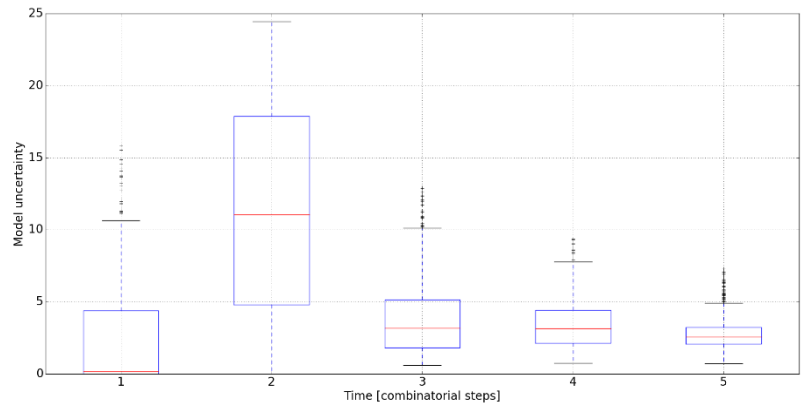

Fig. 3. Model uncertainty reduction over time (the initial high confidence corresponds to neural network initialization parameters)

It is obvious that with time, the uncertainty inherent in the representation decreases. This is accompanied by a similar decay in the prediction error for the transition function. Thus as time passes and the reinforcement agent gathers more experiences, it not only improves its representation but also becomes more confident in its predictions for future states. These reductions in uncertainty can be summarized by the agent exploring new, hitherto unexplored regions of the state-space which we refer to as information gain. The information gain in input feature space, as shown in figure 4, shows experiences for different agents over 25 time steps with each time step corresponding to a period of two days. The boxplot indicates that initially the agents quickly gather experience as they have no prior knowledge of the storage vessel, this however tapers off rapidly as most new interactions with the system lead to states already observed by the agent (this is caused by the fact that human behaviour is periodic to a certain degree).

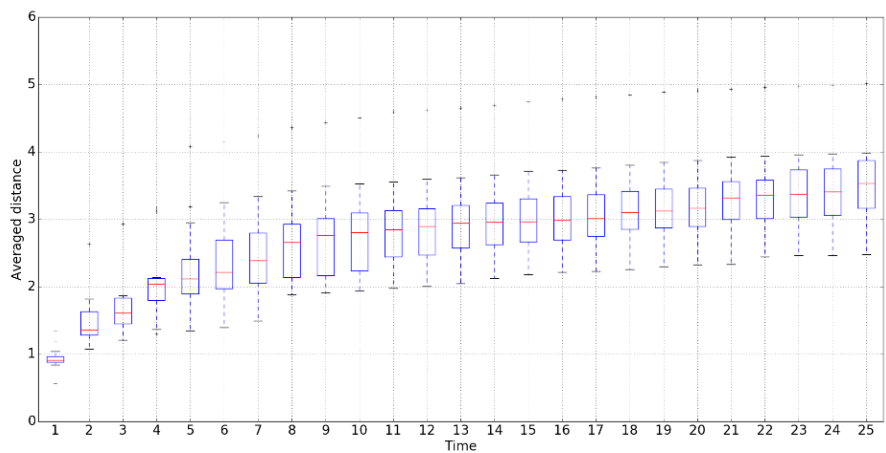

Fig. 4. Information gain as a function of time

This is not to say that each agent has explored the entire state-space; far from it. Results from uncertainty in the output space (figs. 2 and 3) show that the representations learnt by the agents are still fairly uncertain, and while they continue to improve with time this information gain understandably decelerates over time.

2. Offline validation of storage vessel model

To investigate the accuracy and generalization potential of the learnt vessel test, we carried out specific offline tests on a lab setup designed to emulate real world houses. Fig. 5 summarizes some results from these tests where different consumption profiles were used to compare the model predictions and the sensor observations. At the end, hot water was tapped off from the storage vessel and its temperature recorded. This was then compared with model predictions that were generated in a way similar to the plots from Fig. 2.

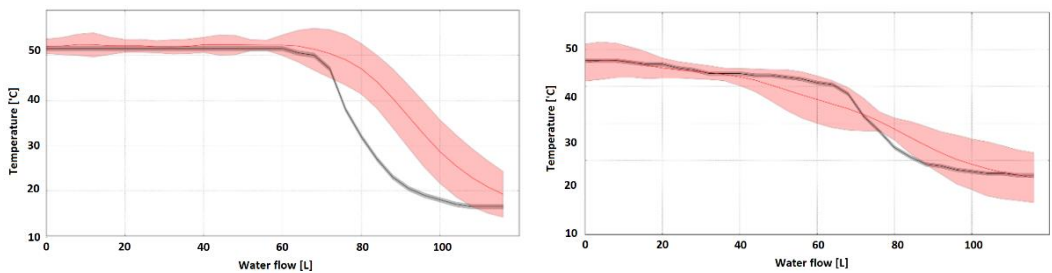

Fig. 5. Offline model evaluation by comparing sensor data (black curve) with model prediction (red curve) for different consumption profiles: (a) tapping off hot water immediately after a reheat cycle; (b) tapping off hot water after 6 hours of intermittent consumption by the user

Tests shows that the algorithm was able to learn the vessel behaviour for different consumption profiles, especially once the uncertainty of the models was taken into account. Model performance generally deteriorated where huge amounts of water were suddenly drawn from the vessel in a way that would impact the occupant comfort. This makes intuitive sense since the heat pump always reheats the storage vessel before the reinforcement agent observes these states and therefore generalization here remains poor. Over time and as the agent gathers more experiences (the learning progress evidenced in Figs. 3 and 4) this performance will further improve. It has to be noted however that, considering the predictive uncertainty, the scale of error in our experiments on an absolute scale was around 10 litres and always less than 15% for regions of interest in the state space. This error can be easily incorporated by making risk-averse predictions.

3. Online validation of storage vessel model

During physical operation of the storage vessel, a mid-point sensor is used to serve as validation (for unseen data) against a vessel model that has been trained on historic data obtained from the same sensor. Fig. 6 plots the observations against the predictions, with the shaded bars representing the uncertainty bounds around the prediction. In this figure, the model was only applied to the discharging period during successive reheat cycles, therefore the behaviour of the model during the charging cycle is undefined (thus explaining the large gap between prediction and measurement). As the duration of the charging cycle is quite short (typically less than an hour) this does not significantly impact the subsequent optimal control step. Furthermore, the agent has no control of vessel state during this time as the heat pump under consideration does not allow modulation of power as a control variable.

The error during the discharge cycle between observation and prediction is usually less than 0.5°C as shown in Fig. 6, which is close to the sensor's tolerance. The prediction almost always remains within the confidence bounds returned by the model. As evident, the model has learnt to predict sudden temperature drops accompanying water consumption by the user. As highlighted before, the uncertainty surrounding these events is usually higher than more frequently observed states. However the next time the agent encounters a similar state, its estimation and prediction capabilities improves because of this experience.

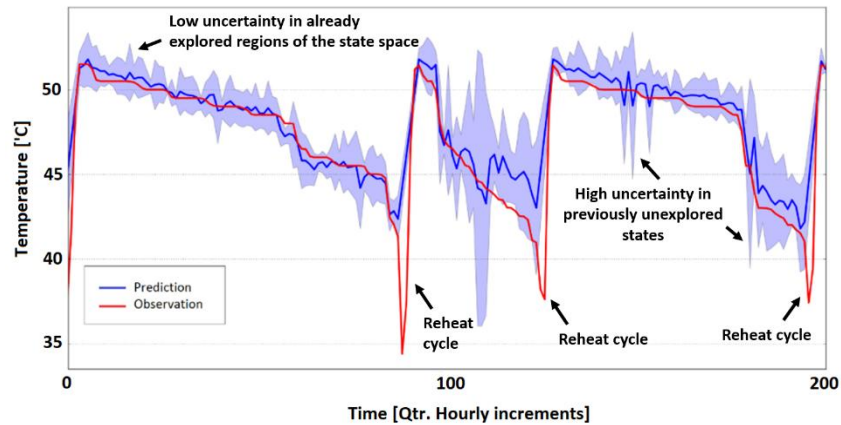

Fig. 6. Online model evaluation: predicted vs. observed water temperature in the storage vessel

4. Heat pump model

The behaviour of the heat pump when reheating the storage vessel can be derived either thermodynamically or from the manufacturer specification sheets. Thermodynamic calculations require information on mass flow rates etc. that are not available generally. Likewise, specification sheets usually given in the form of tables as a function of ambient and inlet water temperature are not available for most existing systems. Furthermore, the COP given in these tables is only indicative and planning with it might lead to consistently suboptimal policies.

We posit that the heat pump model, like the storage vessel model, can be learnt from data. The energy consumption of the heat pump is a strong function of the state of the vessel and the ambient conditions. The vessel state representation learnt earlier can be used for this purpose while (forecasts for) ambient conditions are usually readily accessible from the internet. The heat pump model derived in this way can be used to calculate the energy efficiency rewards for the reinforcement agent. In our experiments, the heat pump behaviour is fairly (piece-wise) linear. The mean absolute error (MAE) obtained while fitting the heat pump model is between 100 and 150 Wh and is normally distributed. This corresponds to a mean absolute percentage error (MAPE) of less than 10% which makes it well suited for use in the reinforcement framework.

5. Occupant models

Occupant behaviour can't be learnt in a manner similar to the storage vessel because of the inherent stochasticity of human behaviour. Furthermore, all the variables relating to human behaviour are latent in the sense that, owing to limited sensing, reliable occupancy estimation is not possible. To make predictions about occupant behaviour, we follow the approach presented in (Kazmi H. e., 2016).

**Energy efficiency gains**

The energy required by the heat pump to reheat the storage vessel is essentially a function of three variables (and their interactions). These include (1) thermodynamic losses (a property of the storage vessel), (2) hot water consumption (a property of occupant behaviour) and (3) ambient conditions (a property of the environment and how it affects the heat pump).

For this experiment, the first and third variables were fixed because of the nature of the experiment whereby control was running on similar heat pumps simultaneously in the same geographical location. Occupant behaviour however remained an uncontrollable influence as different houses consume different amounts of hot water each day. We have tried to minimize the effect of large disparities in water flow across different houses by considering houses with similar hot water demand and comparing their energy consumption first to allow the differences to come forth. The weekly water and energy consumption for these 10 similar houses is visualized in Fig. 7.

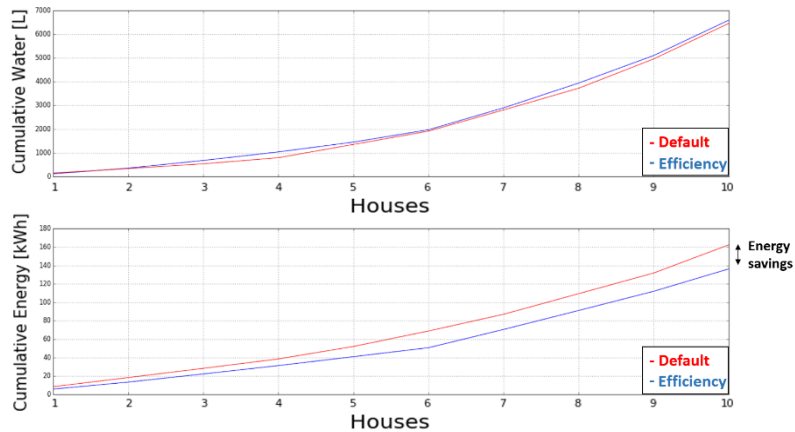

Fig. 7. Water and energy consumption across households running the default and energy efficiency controllers

It is evident from the two plots that while the hot water demand grows fairly comparably for both groups, the energy consumption remains substantially lower for the efficiency group by about 20%. We replicated the experiment over three to four weeks and obtained similar results. These results are summarized in Fig. 8 which plots the daily water demand with daily energy consumption and shows how the two control strategies compare with each other.

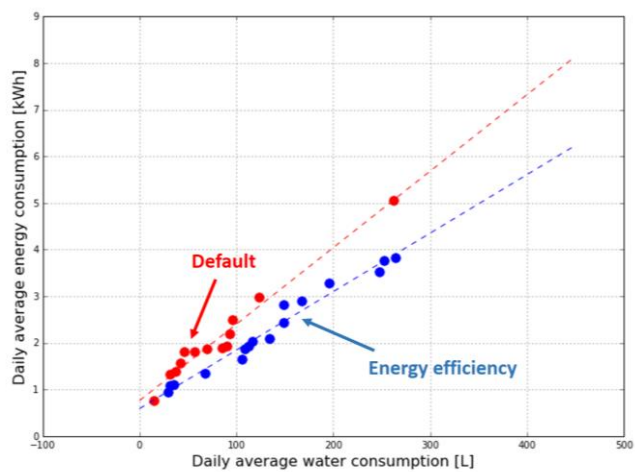

Fig. 8. Energy and water consumption for houses in the different control groups

These results provide further justification for the earlier energy savings where we controlled for the hot water demand. Two fundamental questions remain, how are these savings realized in practice and what is their impact on occupant comfort. To answer the first question, consider Fig. 9 which breaks down the temperature sensor readings time series into a set of discrete episodes. Each episode starts when the storage vessel is reheated and terminates with the initiation of the next reheat cycle. This is done in a similar way as Fig. 2, only these temperature readings come from sensors installed in real houses.

1. The **default reheat cycles** show a uniform, predictable behaviour where the storage vessel is reheated to a certain temperature whereupon it starts losing energy manifested as a drop in mid-point temperature (both due to ambient losses and hot water draws). This culminates in reheat cycles which mostly occur at a time between 10 and 15 hours when the temperature in the vessel has hit the specified threshold of 47.5°C.
2. The **energy efficient controller** drives the temperature in the storage vessel lower, sometimes almost approaching 40°C. This appears to be out of the prescribed comfort bounds but this is the mid-point temperature and not the temperature the occupant is consuming hot water at. The impact of this behaviour on occupant comfort is considered next. The time between average cycles can also be longer for the case of the energy efficient controller, approaching 20 hours.

Another thing of interest in Fig. 9 is the legionella cycles; these are usually carried out by heating the vessel up to 65°C every two weeks and is unchanged between the default and energy efficiency controller. While this too can be brought under the ambit of optimal control, it was considered prudent to not alter this health related aspect of the hot water system.

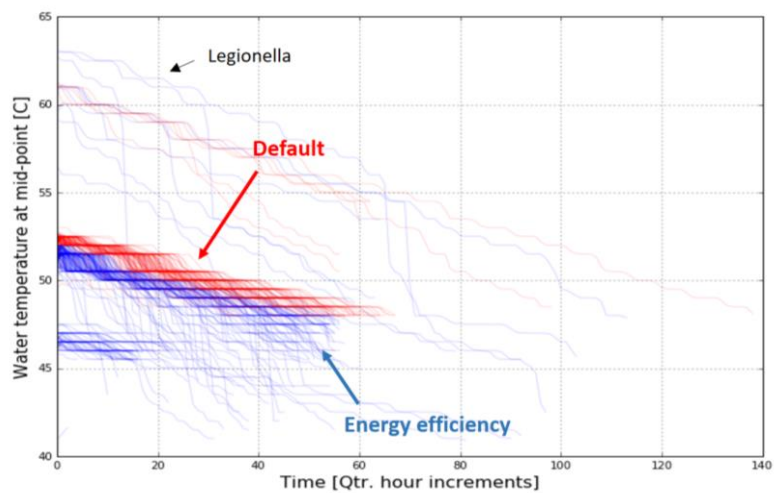

Fig. 9. Vessel behaviour for energy efficiency and default controllers; the default controller exhibits predictable behaviour because of its rule based nature while the energy efficient controller drives the storage vessel to lower temperatures by making use of its knowledge of remaining energy content in the storage vessel

**Impact on occupant comfort**

As defined above, 45°C at the outflow is the lower bound to maintain occupant comfort. Any hot water draws below this temperature are considered to have violated occupant comfort. Since Fig. 9 visualizes water temperature at the mid-point and not at the outflow, we consider the five houses specifically fitted with temperature sensors at the outflow for occupant comfort. This is done to validate that occupant comfort has not been violated while improving energy efficiency. These results are shown in Fig. 10.

It is evident in Fig. 10 that the outflow temperature doesn't fall below 45 degrees for any of the five houses with additional sensors. This provides documentary evidence that energy efficiency isn't at the cost of occupant comfort which is the primary objective of this research. It also brings to light the fact that the reinforcement agent has successfully managed to postpone consumption in time (by letting temperature drop further than in the default case) in such a way that it doesn't negatively impact occupant comfort.

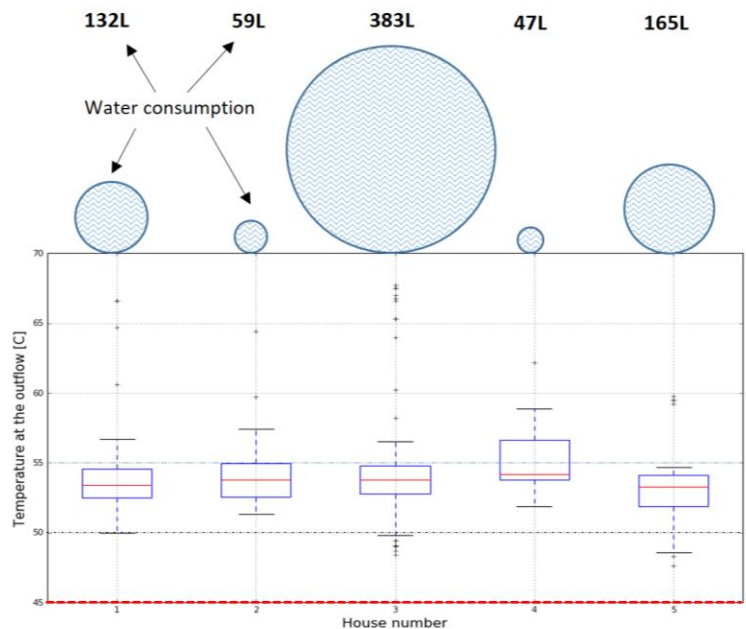

Fig. 10. The impact of efficiency controller on occupant comfort and its link with hot water demand

Fig. 10 also makes it possible to explore the link between occupant comfort and hot water consumption. The probability of lower outflow temperature rises with higher hot water demand and vice versa. This is especially obvious for the three classes the houses can be distinguished into. House 2 and 4 exhibit low demand, house 1 and 5 medium demand and house 3 shows unusually high demand. While the median temperature is very similar for all the houses (at around 53°C), the spread is quite different. For the low demand houses, the minimum outflow temperature doesn't drop below 50°C at any time during the observation period, while for the medium and high demand houses it sometimes drops below 50°C (but still remains above the 45°C threshold).

The reason for this difference is the backup controller being employed to ensure hot water supply at all times. For the low demand houses, this frequently comes into play because while the temperature at the top of the vessel might be above the threshold according to the learnt model and sufficient hot water remains in the vessel to meet hot water demand of the occupant, the total amount of hot water in the vessel to cover contingencies has dropped below the threshold forcing a reheat cycle. This is not the case for houses with higher demand because the reheat cycle is usually initiated much earlier irrespective of the backup controller.

**Discussion**

The algorithmic framework developed in this paper retains most of the benefits offered by Model Predictive Controllers without suffering from the most important modelling drawback. By being completely data-driven, the algorithm requires no offline computations or human involvement. In this plug-and-play approach, it can be integrated with the millions of hot water devices in residential and commercial use throughout the world. The only sensor requirements are for temperature in the storage vessel, a hot water flow meter and energy measurement. These sensors are already available by default in many situations. In the future, as sensing technologies grow more ubiquitous, the availability of data will only grow further facilitating such control algorithms.

While this work has demonstrated the framework on a particular hot water system, it has to be emphasized that the learning algorithm is agnostic to the type of storage vessel or the heating equipment it integrates with. Likewise, weather and occupant influences do not pose a problem for generalization as they are learnt from data. If one of these characteristics changes, e.g. a household replaces the storage vessel, the reinforcement agent will learn to adapt to the behaviour of this new storage vessel. This is opposed to the case of MPC where sub-optimal policies would continue to be executed because of an incorrect dynamics model.

In addition to different hot water systems, different optimization objectives can also be seamlessly handled with the proposed methodology. An example of a different optimization objective function would be mitigating the potential effects of electrification in existing grids. These include (1) self-consumption of local solar generation because of possible cost differentials (Castillo-Cagigal, 2011), where the hot water vessel is charged when excess solar energy is available and (2) peak shaving where rising grid reinforcement costs (Dupont, 2012) can be reduced by using the storage vessel to provide flexibility to reduce injection or off-take peaks (Koen Vanthournout, 2012) (Kazmi H. a., 2016).

With the proposed reinforcement learning controller, it is straightforward to achieve these different objectives: the occupant comfort and learning components remain unchanged, but the efficiency constraint is replaced (or complemented) by self-consumption or peak shaving. To do so, additional forecasts need to be much such as local solar production and baseload consumption. Furthermore, peak shaving is a coordination problem where centralized planning algorithms can be employed in conjunction with the reinforcement framework highlighted previously. Alternatively, distributed algorithms such as dual decomposition (Xiao, 2004) and ADMM (Boyd, 2011) can be used for the planning phase in case there are data privacy concerns.

Nevertheless, the framework can seamlessly integrate such changes in the objective function. This offers an advantage over model-free reinforcement learning techniques which suffer from two key disadvantages in this regard: (1) a new optimization objective will necessitate a new reward function which will lead to starting policy search over from scratch, and (2) lower sample efficiency, i.e. model free controllers require more data to attain the same performance.

**Conclusions**

In this paper, we have presented a model-based reinforcement learning algorithm to optimize energy efficiency of hot water production. The controller reduced the energy consumption by almost 20% for a set of 32 Dutch houses while maintaining occupant comfort measured in an objective manner. Extrapolated to a year, this has the potential to reduce household energy consumption by up to 200 kWh.

A key benefit of the proposed framework is that it can be seamlessly extended to other thermal systems or to include other objective functions such as maximization of solar consumption or peak shaving etc. The model learnt for optimal control can also be used to provide other valuable services e.g. in a recommender system to inform end user choices or as a monitoring system for heat pump manufacturers to track operational efficiency and compare it to design specifications.

Being completely data-driven, the replication potential of the proposed algorithm is enormous and it could easily lead to gigawatt scale savings in only the European context. With over 200 million households only in Europe, absolute energy savings of hundreds of kWh on an annual basis per household has the potential to reduce energy consumption by tens to hundreds of terra-watt hours. This is on a similar scale as the energy consumption of countries such as Belgium and The Netherlands. More realistically, even a fractional proliferation into a market of this size would lead to annual efficiency improvements of hundreds of gigawatt hours in addition to bringing about the corresponding economic and health benefits. The proposed framework allows us to unlock these efficiency improvements at very little to no extra cost in existing and new hot water systems.

**Acknowledgments**

This work was done with support from Enervalis, VLAIO (formerly IWT) and InnoEnergy. The authors also thank the referees for their constructive input.

**References**

Afram, A. a.-S. (2014). Review of modeling methods for HVAC systems. *Applied Thermal Engineering 67.1*, 507-519.

Afram, A. a.-S. (2014). Review of modeling methods for HVAC systems. *Applied Thermal Engineering 67.1*, 507-519.

Afram, A. a.-S. (2014). Theory and applications of HVAC control systems–A review of model predictive control (MPC). *Building and Environment 72*, 343-355.

Badescu, V. (2007). Optimal control of flow in solar collectors for maximum exergy extraction. *International Journal of Heat and Mass Transfer*, 4311-4322.

Badescu, V. (2008). Optimal control of flow in solar collector systems with fully mixed water storage tanks. *Energy Conversion and Management 49*, 169-184.

Balian, R. (2004). *Entropy, a protean concept.* Poincaré Seminar.

Boyd, S. (2011). Alternating direction method of multipliers. *NIPS Workshop on Optimization and Machine Learning.*

Castillo-Cagigal, M. e. (2011). PV self-consumption optimization with storage and active DSM for the residential sector. *Solar Energy 85.9*, 2338-2348.

Chua, K. J. (2010). Advances in heat pump systems: A review. *Applied energy 87.12*, 3611-3624.

Dane George, N. S. (2015). High resolution measured domestic hot water consumption of Canadian homes. *Energy and buildings*, 304-315.

Deisenroth, M. a. (2011). PILCO: A model-based and data-efficient approach to policy search. *28th International Conference on machine learning.*

Dorigo, M. M. (2006). Ant colony optimization. *IEEE computational intelligence magazine 1.4*, 28-39.

Dupont, B. e. (2012). LINEAR breakthrough project: Large-scale implementation of smart grid technologies in distribution grids. *IEEE Innovative Smart Grid Technologies Europe.*

F. De Ridder, M. D. (2011). An optimal control algorithm for borehole thermal energy storage systems. *Energy Build., vol. 43, no. 10*, 2918–2925.

Frederik Ruelens, B. C. (2014). Demand Response of a Heterogeneous Cluster of Electric Water Heaters using Batch Reinforcement Learning. *18th Power Systems Computation Conference*. Wroclaw.

Frederik Ruelens, S. I. (2015). Learning Agent for a Heat-Pump Thermostat with a Set-Back Strategy Using Model-Free Reinforcement Learning. *Energies*.

Gal, Y. R. (2016). Improving PILCO with bayesian neural network dynamics models. *Data-Efficient Machine Learning workshop. Vol. 951.*

Gill, Z. M. (2010). Low-energy dwellings: the contribution of behaviours to actual performance. *Building Research & Information 38.5*, 491-508.

Hester, T. a. (2013). TEXPLORE: real-time sample-efficient reinforcement learning for robots. *Machine learning 90.3*, 385-429.

IEA. (2016). *Energy efficiency indicators.*

IEA. (2017). *Renewables information: overview.*

IPCC . (2014). *The Fifth Assessment Report of the Intergovernmental Panel on Climate Change.*

IRENA. (2017). *Renewable energy and jobs - annual review.*

James, B. (2003). Load Control Using Building Thermal Mass. *Trans. ASME, vol. 125*.

Jose Vazquez-Canteli, J. K. (2017). Balancing comfort and energy consumption of a heat pump using batch reinforcement learning with fitted Q-iteration. *Energy Procedia 122*, 415-420.

Kazmi, H. a. (2016). Demonstrating model-based reinforcement learning for energy efficiency and demand response using hot water vessels in net-zero energy buildings. *PES Innovative Smart Grid Technologies Conference Europe.* Ljubljana.

Kazmi, H. e. (2016). Generalizable occupant-driven optimization model for domestic hot water production in NZEB. *Applied Energy 175* , 1-15.

Koen Vanthournout, R. D. (2012). A Smart Domestic Hot Water Buffer. *IEEE transactions on smart grids*, 2121 - 2127.

Lawler, E. L. (1966). Branch-and-bound methods: A survey. *Operations research 14.4*, 699-719.

Majcen, D. (2016). *Predicting energy consumption and savings in the housing stock: A performance gap analysis in the Netherlands.*

Masoso, O. T. (2010). The dark side of occupants' behaviour on building energy use. *Energy and buildings 42.2*, 173-177.

Oldewurtel, F. e. (2012). Use of model predictive control and weather forecasts for energy efficient building climate control. *Energy and Buildings 45*, 15-27.

Pérez-Lombard, L. O. (2008). A review on buildings energy consumption information. *Energy and buildings, 40(3)*, 394-398.

Santin, O. G. (2009). The effect of occupancy and building characteristics on energy use for space and water heating in Dutch residential stock. *Energy and buildings 41.11*, 1223-1232.

Srinivas, N. a. (n.d.). Muiltiobjective optimization using nondominated sorting in genetic algorithms. *Evolutionary computation 2.3*, 221-248.

Tianshu Wei, Y. W. (2017). Deep Reinforcement Learning for Building HVAC Control. *54th Annual Design Automation Conference .*

Weather Underground. (2017, 11 10). *Weather Underground API*. Retrieved from https://www.wunderground.com/weather/api/

Xiao, L. M. (2004). Simultaneous routing and resource allocation via dual decomposition. *IEEE Transactions on Communications 52.7*, 1136-1144.

Zhang, J. &. (2007). Best switching time of hot water cylinder-switched optimal control approach. *AFRICON* (pp. pp. 1-7). IEEE.